\documentclass[prl,twocolumn,amsmath,amssymb,superscriptaddress,showpacs]{revtex4}
\usepackage{graphics}
\usepackage{epsfig}
\usepackage{bbm}
\usepackage[latin1]{inputen c} % Umlaute!

\newcommand{\be}{\begin{equation}}
\newcommand{\ee}{\end{equation}}
\newcommand{\eea}{\end{eqnarray}}

\newcommand{\exs}[1]{\ensuremath{\langle{#1}\rangle}}
\newcommand{\mean}[1]{\ensuremath{\langle{#1}\rangle}}

\newcommand{\eins}{\openone}

\newcommand{\ketbra}[1]{\ensuremath{| #1 \rangle \langle #1 |}}

\newcommand{\ket}[1]{\ensuremath{|#1\rangle}}

\newcommand{\bra}[1]{\ensuremath{\langle#1|}}

\newcommand{\kommentar}[1]{}
\newcommand{\trace}{{\rm Tr}}

\renewcommand{\vr}{\varrho}
\newcommand{\FF}{\ensuremath{\mathcal{F}}}

\begin{document}
\title{
Entanglement and Permutational Symmetry
}
\date{\today}
\begin{abstract}
We study the separability of permutationally
symmetric quantum states. We show that for bipartite symmetric
systems most of the relevant entanglement criteria coincide.
However, we provide a method to generate examples of bound
entangled states in symmetric systems, for the bipartite and
the multipartite case. These states shed some new light on the
nature of bound entanglement.

\end{abstract}

\author{G\'eza T\'oth}
%\email{toth@alumni.nd.edu}
\affiliation{Department of Theoretical Physics, The University of the Basque Country, P.O. Box 644, E-48080 Bilbao, Spain} \affiliation{Ikerbasque-Basque Foundation for
Science, Alameda Urquijo 36,
E-48011 Bilbao, Spain}
\affiliation{ICFO-The Institute of Photonic Sciences, E-08860
Castelldefels (Barcelona), Spain}
\affiliation{Research Institute for Solid State Physics and Optics,
Hungarian Academy of Sciences, \\ P.O. Box 49, H-1525 Budapest,
Hungary}

\author{Otfried G\"uhne}
%\email{otfried.guehne@uibk.ac.at}
\affiliation{Institut f\"ur
Quantenoptik und Quanteninformation, \"Osterreichische Akademie der
Wissenschaften,\\ A-6020 Innsbruck, Austria} \affiliation{Institut
f\"ur Theoretische Physik, Universit\"at Innsbruck,
Technikerstra{\ss}e 25, A-6020 Innsbruck, Austria}


\maketitle

% 03.65.Ud                  Entanglement and quantum nonlocality
% 03.67.Mn                  Entanglement in Quantum Information
% 03.67.-a                  Quantum information
% 03.67.Lx                  Quantum computation
% 02.70.-c                  Computational techniques

%%%%%%%%%%%%%%%%%%%%%%%%%%%%%%%%%%%%%%%%%%%%%%%%%%%%%%%%%%%%%%%%%%%%%%

%\section{Introduction}

Entanglement is a central phenomenon of quantum mechanics
and plays a key role in quantum information processing
applications such as quantum teleportation and quantum cryptography
\cite{Hreview}. Therefore, entanglement appears as a natural goal
of many recent experiments aiming to create various quantum states
with photons, trapped ions, or cold atoms in optical lattices.
While being at the center of attention, to decide whether a quantum
state is entangled or not is still an unsolved problem. There are numerous
criteria for the detection of entanglement, but no general solution
has been found \cite{Hreview}.

Symmetry is another central concept in quantum mechanics \cite{symmetry}.
Typically, the presence of certain symmetries simplifies the
solution of tasks like the calculation of atomic spectra or
finding the ground state of a given spin model. Symmetries are also useful
in quantum information theory: For instance,
if a multiparticle quantum state is invariant un der the same change
of the basis at all parties (i.e., invariant under local
unitary transformations of the type
$U_{\rm tot} = U \otimes U \otimes ... \otimes U$), this symmetry
can be used to study the existence of local hidden variable models
\cite{W89}, to determine its entanglement properties \cite{eggeling} or
to simplify the calculation of entanglement measures
\cite{vollbrecht}.

In this Letter, we investigate to which extent symmetry under
permutation of the particles simplifies the separability problem.
In general, a quantum state $\vr$ is called separable, if it can be written as
%mixture of product states
%\be
$\vr= \sum_k p_k \vr^A_k \otimes\vr^B_k,$
%\ee
where the $p_k$ form a probability distribution. There
are several necessary criteria for a state to be separable.
The most famous one is the criterion of the positivity of
the partial transposition (PPT), which states that for a
separable
$\vr = \sum_{ij,kl} \vr_{ij,kl} \ket{i}\bra{j}\otimes \ket{k}\bra{l}$
the partially transposed state
$\vr^{T_A}=\sum_{ij,kl} \vr_{ji,kl} \ket{i}\bra{j}\otimes\ket{k}\bra{l}$
has no negative eigenvalues \cite{ppt}. This criterion is necessary
and sufficient only for small systems ($2\times 2$ and $2\times 3),$
while for other dimensions some entangled states escape the detection
\cite{bound}. These states are then  bound entangled, which means that
no pure state entanglement can be distilled from them. While bound
entangled states are difficult to construct, they play an important
role in quantum information theory, as they are at the heart of some
open problems  in quantum information theory \cite{bound, boundinteresting}.
Apart from the PPT criterion, several other strong separability criteria
exist \cite{ccnr,ccnr2,GH06,zhang,DP04,B06}, which can detect
some states where the PPT criterion fails. Also for symmetric
states, some special separability criteria have been proposed
\cite{corrsym,KC05,ES02,wangsanders}.

We will show that for states that are symmetric under a permutation
of the particles, most of the relevant known separability criteria coincide.
However, we present examples of bound entangled symmetric states.
These states form therefore a challenge for the derivation of new
separability criteria. Moreover, these states shed new light on the
phenomenon of bound entanglement, as it has been suggested
that symmetry and bound entanglement are contradicting
notions \cite{H03,HH06,fannes}. Finally, we present symmetric
multipartite bound entangled states, which are nevertheless genuine
multipartite entangled.

We first consider two $d$-dimensional quantum systems. We
examine two types
of permutational symmetries, denoting the corresponding sets by
$\mathcal{I}$ and $\mathcal{S}:$

(i) We call a state {\it permutationally invariant}
(or just invariant, $\vr \in \mathcal{I}$) if $\vr$
is invariant under exchanging the particles. This can be formalized
by using the flip operator $F = \sum_{ij}\ket{ij}\bra{ji}$ as
$F \vr F = \vr.$ The reduced state of two randomly chosen particles
of a larger ensemble has this symmetry.

(ii) We call a state {\it symmetric} ($\vr\in \mathcal{S}$) if
it acts on the symmetric subspace only. This space is spanned
by the basis vectors
$\ket{\phi_{kl}^{+}}:=(\ket{k}\ket{l}+\ket{l}\ket{k})/\sqrt{2}$ for
$k\ne l$ and $\ket{\psi_k}:=\ket{k}\ket{k}.$ The projector $P_s$
onto this space can be written as $P_s=(\eins + F)/2.$ This implies
that for symmetric states by definition $P_s \vr P_s =\vr P_s = \vr$
and $\vr F = F \vr = \vr.$ This is the state space of two $d$-state
bosons.

Clearly, we have $\mathcal{S} \subset \mathcal{I}.$ For a
basis state of the antisymmetric subspace
$\ket{\phi_{kl}^{-}}:=(\ket{k}\ket{l}-\ket{l}\ket{k})/\sqrt{2}$,
we have, $\ketbra{\phi_{kl}^{-}} \in  \mathcal{I},$
however, $\ketbra{\phi_{kl}^{-}} \notin  \mathcal{S}.$
Our main tool for the investigation of entanglement criteria
is an expectation value matrix of a bipartite quantum state.
This matrix has the entries
\begin{equation}
\eta_{kl}(\vr):=\exs{M_k \otimes M_l}_\vr,\label{eta}
\end{equation}
where $M_k$'s are local orthogonal observables for both parties,
satisfying ${\rm Tr}(M_k M_l)=\delta_{kl}$ \cite{moroder}.
We can directly  formulate our first main result:
\\
{\bf Observation 1. } Let $\vr\in \mathcal{S}$ be a symmetric
state. Then the following separability criteria are equivalent:
\\
(i) $\eta \geq 0,$ or, equivalently $\mean{A \otimes A } \geq 0$ for all observables $A$ \cite{wangsanders}.
\\
(ii) $\vr$ has a positive partial transpose, $\vr^{T_A} \geq 0$
\cite{ppt}.
\\
(iii) $\vr$ satisfies the CCNR criterion, $\Vert R(\vr)\Vert_1 \leq 1,$
where $R(\vr)$ denotes the realignment map and $\Vert ... \Vert_1$
denotes the trace norm \cite{ccnr,ccnr2}.
\\
(iv) The correlation matrix, defined via the matrix elements as
\begin{equation}
C_{kl}:=\exs{M_k \otimes M_l}-\exs{M_k \otimes \mathbbm{1}}\exs{\mathbbm{1} \otimes M_l}
\end{equation}
is positive semidefinite \cite{corrsym}.
\\
(v) The state satisfies several variants of the covariance matrix
criterion, e.g., $\Vert C \Vert^2_1 \leq
[1-\trace(\vr_A^2)][1-\trace(\vr_B^2)]$ or $ 2 \sum |C_{ii}|
\leq [1-\trace(\vr_A^2)]+[1-\trace(\vr_B^2)]$ \cite{GH06,zhang}.
These criteria are for general states strictly stronger
than the CCNR criterion.

Here, separability criteria are formulated as conditions that
a separable state has to fulfill, and violation implies entanglement
of the state. Further, we call two separability criteria
equivalent, if a state that is detected by the first criterion
is also detected by the second one and vice versa. Note that
the criteria (i) and (iv) are criteria specifically for symmetric
states, while the others also work for generic separable states.
During the proof of this theorem, we will see that several of the
equivalences also hold for permutationally invariant states.

{\it Proof.}
For invariant states, $\eta$ is a real symmetric matrix.
It can be diagonalized by an orthogonal
matrix $O.$ The resulting diagonal matrix $\{\Lambda_k\}$
is the correlation matrix corresponding to the observables
$M_k'=\sum O_{kl} M_l.$ Hence, any invariant state
can be written as
\begin{equation}
\vr=\sum_k \Lambda_k M'_k \otimes M'_k, \label{Schdecomp2}
\end{equation}
where $M_k'$ are pairwise orthogonal observables. This is almost the
Schmidt decomposition of the matrix $\vr,$ with the only difference
that $\Lambda_k$ (which are the eigenvalues of $\eta$) can also be negative.
Let us compute $\trace(\eta)=\sum_k \Lambda_k=\exs{\sum_k M_k'
\otimes M'_k}$. We can use that
$ \sum_k M_k' \otimes M'_k = F, $ where $F$ is again the
flip operator \cite{orthog}. Hence, $ -1\le\sum_k \Lambda_k \le 1$
for invariant states and $\sum_k \Lambda_k=1$ for symmetric
states.

Now we can show the first equivalences. Let us start with the CCNR
criterion. It states that if $\vr$ is separable, then $\Vert
R(\vr)\Vert_1 \leq 1$ where $\Vert X \Vert_1 = {\rm Tr}(\sqrt{X
X^\dagger})$ is the trace norm and the realigned density matrix is
$R(\vr_{ij,kl}) = \vr_{ik,jl}$ \cite{ccnr2}. As noted in
Ref.~\cite{Mth}, this just means that if $\|(\vr F)^{T_A}\|_1 > 1$ then
$\vr$ is entangled. Since for symmetric states $\vr F=\vr,$ this
condition is equivalent to $\|\vr^{T_A}\|_1>1.$ This is just the
PPT criterion, since we have ${\rm Tr}(\vr^{T_A})=1$ and the condition
$\|\vr^{T_A}\|_1>1$ signals the presence of negative eigenvalues.

On the other hand, the realignment criterion can be reformulated
as follows: If $\sum_k \vert\Lambda_k\vert > 1$ in the Schmidt decomposition
[Eq.~(\ref{Schdecomp2})], then the quantum state is entangled
\cite{ccnr}. For symmetric states this is equivalent to $\Lambda_k
< 0$ for some $k.$ But then $\mean{M_k'\otimes M_k'}<0$ and $\eta$ has
a negative eigenvalue. For invariant states $\Lambda_k <0$ is
necessary but not sufficient for violating the CCNR criterion. This
proves the equivalence of (i),(ii) and (iii).

Now we show that $C\ge 0 \Leftrightarrow \eta\ge 0.$ The direction
``$\Rightarrow$'' is trivial, since for invariant states the
matrix $\exs{M_k \otimes \openone} \exs{\openone\otimes M_l}$ is a
projector and hence positive. On the other hand, if we make for a
given state the special choice of observables $Q_k= M_k - \mean{M_k}$
we just have $C(M_k)=\eta(Q_k),$ implying the other direction. Here we use that
$\eta(M_k)\ge 0$ implies that $\eta(Q_k)\ge 0$
for arbitrary (not pairwise orthogonal) observables $Q_k.$
Note
that here the invariance of the state guaranteed that the $Q_k$
are the same for both parties.

Finally, let us turn to the discussion of (v). If $\vr$ is
symmetric, the fact that $C$ is positive semidefinite gives
$\Vert C \Vert_1 = \trace(C) = \sum \Lambda_k - \sum_k
\trace(\vr_A M_k')^2 = 1 - \trace(\vr_A^2)$ and similarly, $\sum_i
|C_{ii}| = \sum_i C_{ii} = 1 - \trace(\vr_A^2).$ Hence, a
state fulfilling (iv) fulfills also both criteria in (v). On the
other hand, a state violating (iv) must also violate the conditions
in (v), as they are strictly stronger than the CCNR criterion
\cite{basisremark}.
$\hfill \blacksquare$

In this proof, sometimes only the permutational invariance
was used, hence we can state for invariant states:\\
{\bf Observation 2.} For invariant states, the separability
criteria (i), (iv) and (v) are equivalent.

Given the equivalence of the above criteria, it is an
interesting question to ask whether there are any
entangled symmetric states that escape the detection
by these criteria. These states are then PPT and hence
bound entangled. Moreover, such states are a challenge
for the derivation of new separability criteria, as most of
the standard criteria fail. We will present now two methods
for constructing such states.

The literature has already examples for invariant bound
entangled states:
In Ref.~\cite{B06} Breuer presented, for  even $d \ge 4$, a single parameter
family of bound entangled states that are $\mathcal{I}$ symmetric, namely
\begin{equation}
\vr_{\rm B}=\lambda \ketbra{\Psi_{0}^d} + (1-\lambda) \Pi_s^d.
\label{breuer}
\end{equation}
The state is shown to be entangled for $0<\lambda\le1$ while it is
PPT for $0\le\lambda\le1/(d+2).$ Here $\ket{\Psi_0}$ is the singlet
state [for the $d=4$ case it is
$\ket{\Psi_{0}}=(\ket{03}-\ket{12}+\ket{21}-\ket{30})/2$] and $\Pi_s$
is the normalized projector to the symmetric subspace,
$\Pi_s^d = P_s /[d(d+1)/2]$ \cite{wolfetal}.

The first idea to construct bound entangled states with $\mathcal{I}$-
or $\mathcal{S}$-symmetry is to embed a low dimensional entangled state
into a higher dimensional Hilbert space, such that it becomes symmetric,
while it remains entangled. To see a first example, consider a general
bound entangled state $\vr_{AB}$ on $\mathcal{H}_{AB}$ and add to each
party a two-dimensional Hilbert space $\mathcal{H}_{A'}$ and
$\mathcal{H}_{B'}.$
Then one can consider the state
\begin{equation}
\tilde{\vr} =
\frac{1}{2}
[\ketbra{10}_{A'B'}\otimes \vr_{AB}
+\ketbra{01}_{A'B'}\otimes (F\vr_{AB} F)]. \label{rho2}
\end{equation}
If $\vr$ acts on a $d\times d$ system then $\tilde{\vr}$ acts
on a system of size $2d\times 2d.$  Obviously, $\tilde{\vr}$ is
an invariant state. If $\vr$ is entangled, then $\tilde{\vr}$
is entangled, too, as one can obtain $\vr$ from $\tilde{\vr}$
by a local measurement on the ancilla qubits $A'$ and $B'$.
Moreover, if $\vr$ is PPT then $\tilde{\vr}$ is also PPT.
Substituting the various non-symmetric bound entangled states
available in the literature for $\vr$ in Eq.~(\ref{rho2}) gives
invariant bound entangled states.

With a similar method, one can generate a symmetric bound entangled
state. Starting from Eq.~(\ref{breuer}) we consider the state
\begin{equation}
\hat{\vr} =
\lambda \Pi_a^D \otimes \ketbra{\Psi_{0}^d} + (1-\lambda)
\Pi_s^D \otimes \Pi_s^d.
\label{breuer2}
\end{equation}
Here, $\Pi_a^D$ and $\Pi_s^D$ are appropriately normalized projectors to
the two-qudit symmetric/antisymmetric subspace with dimension $D,$
e.g., $\Pi_a^D = P_a/[D(D-1)/2]$ with $P_a = \eins - P_s.$ This
guarantees that  $\hat{\vr}$ is symmetric. Again, if the original
system is of dimension $d\times d$ then the system of $\hat{\vr}$
is of dimension $dD \times dD.$ Since $\vr_{\rm B}$ is the reduced
state of $\hat{\vr},$ if the first is entangled, then the second is
also entangled. However, it is not clear from the beginning that if
$\vr_{\rm B}$ is PPT then $\hat{\vr}$ is also PPT, since $\Pi_a^D$
is entangled. For $D=2$ and $N=4,$ however, numerical calculation
shows that $\hat{\vr}$ is PPT for $\lambda<0.062.$ This provides an
example of an $\mathcal{S}$ symmetric (and invariant) bound entangled
state of size $8\times 8.$
Note that this state represents an explicit counterexample to
Ref.~\cite{fannes}. There, it has been suggested that an
invariant state with $\eta \geq 0$ has to be of the form
$\vr= \sum_k p_k \vr_k \otimes \vr_k,$ which implies that it
is separable \cite{fannesremark}. Also, in Ref.~\cite{H03}
it has been found that the non-distillability of entangled
quantum states may be connected to the asymmetry of quantum correlations
in that state. While this may be valid for many examples,
our results demonstrate that there is not a strict rule
connecting the two phenomena.

Finally, we show a simple method for constructing symmetric
bipartite bound entangled states numerically. We first generate
an $N$-qubit symmetric state, that is, a state of the symmetric
subspace. We consider even $N.$ It is known that such a state is either separable
with respect to all bipartitions or it is entangled with respect
to all bipartitions \cite{newsymmpaper}. Thus any state that is PPT with
respect to the $\tfrac{N}{2}:\tfrac{N}{2}$ partition while NPT with
respect to some other partition is bound entangled with respect to
the $\tfrac{N}{2}:\tfrac{N}{2}$ partition. Since the state is symmetric,
it can straightforwardly be mapped to a $(\tfrac{N}{2}+1) \times (\tfrac{N}{2}+1)$
bipartite
symmetric state \cite{mapremark}.

To obtain such a multiqubit state,  one has to first generate an
initial random state $\varrho$ that is PPT with respect to the
$\tfrac{N}{2}:\tfrac{N}{2}$ partition. Ref.~\cite{ZS01} describes
how to get a random density matrix with a uniform distribution
according to the Hilbert-Schmidt measure. Then, we compute the
minimum nonzero eigenvalue of the partial transpose of $\varrho$
with respect to all other partitions
$
\lambda_{\min}
(\varrho) := \min_k \min_l \lambda_l (\vr^{T_{I_k}}).
$
Here $I_k$ describes which qubits to transpose for the partition
$k.$ If $\lambda_{\min} (\varrho) < 0$ then the state is bound
entangled with respect to the $\tfrac{N}{2}:\tfrac{N}{2}$ partition.
If it is non-negative then we start an optimization process for
decreasing this quantity. The zero eigenvalues due to the nonmaximal rank of
symmetric states are excluded from the minimization,
otherwise one always gets $\lambda_{\min}\le 0.$

We generate another random density matrix $\Delta\vr,$ and
check the properties of
$\varrho'=(1-\varepsilon)\varrho+\varepsilon\Delta\varrho,$
where $0< \varepsilon<1$ is a small constant.
If
$\varrho'$ is still PPT with respect to the
$\tfrac{N}{2}:\tfrac{N}{2}$ partition and $\lambda_{\min}
(\varrho')<\lambda_{\min} (\varrho)$ then we use $\varrho'$
as our new random initial state $\varrho$. If this is not the case, we keep
the original $\varrho.$ Repeating this procedure,  we obtained a
four-qubit symmetric state this way
\be \varrho_{BE4} = {\rm diag}(0.22,0.176,0.167,0.254,0.183)-0.059R,
\nonumber
\ee
where
$R:=\ket{3}\bra{0}+\ket{0}\bra{3}.$ The basis states are
$\ket{0}:=\ket{0000},$ $\ket{1}:={\rm sym}(\ket{1000}),$
$\ket{2}:={\rm sym}(\ket{1100}), ... $  where ${\rm sym}(A)$ denotes
an equal superposition of all permutations of $A$. The state is
bound entangled with respect to the $2:2$ partition. This
corresponds to a $3\times 3$ bipartite symmetric bound entangled state
\cite{mapremark}, demonstrating the simplest possible symmetric bound
entangled state.

Our method can be straightforwardly generalized to create
multipartite bound entangled states of the symmetric subspace,
such that {\it all} bipartitions are PPT (``fully PPT states'').
Then, however, a new separability criterion must be used, different
from the PPT criterion. The PPT symmetric extension of
A.~Doherty {\it et al.} \cite{DP04}
seems to be ideal for our case. For symmetric states, it can be
formulated as follows: We define the PPT symmetric extension of
an $N-$qubit state $\vr_N$ as a symmetric
$M-$qubit state, $\vr_M$ such that
%\be
$\vr_N =
\trace_{N+1,N+2,...,M}(\vr_M),$ %\ee
and all bipartitions of $\vr_M$
are PPT. If
there is an $M$ for which such an extension does not exist then our
state is entangled. Semi-definite programming makes it possible to
look for such an extension. Note that the two density matrices can be efficiently
stored as $(M+1)\times(M+1)$ and $(N+1)\times(N+1)$ matrices,
respectively, in the symmetric basis, making it possible to look for
very large extensions or examine large states \cite{note}.
Moreover, similarly to the algorithm described in the previous
paragraph, it is possible to design a simple random search that,
starting from fully PPT random non-entangled states, leads to
PPT states without an extension. We found such a state for five qubits
\be
\varrho_{BE5} = {\rm diag}(0.17,0.174,0.153,0.182,0.147,0.174)-Q,
\nonumber
\ee where
$Q:=0.0137(\ket{4}\bra{0}+\ket{0}\bra{4}),$ and the basis states of
the symmetric system are $\ket{0},\ket{1},...,\ket{4}.$ We also
found such a state for six qubits \cite{tobepub}.

These multi-qubit states are by construction genuine
multipartite entangled \cite{AB01,newsymmpaper}. This finding is quite peculiar: Genuine
multipartite entanglement is considered in a sense a strong type
of entanglement, while local states or states with PPT bipartitions
are considered weakly entangled.
It is interesting to relate this to the Peres conjecture,
stating that fully PPT states cannot violate a Bell inequality
\cite{P99}. If this is true, then we presented genuine multi-qubit
states that are local. So far, such states have been known
only for the three-qubit case \cite{TA06}.

In summary, we have discussed entanglement in symmetric
systems. We showed that for states that are in the symmetric
subspace several relevant entanglement conditions, especially the
PPT criterion, the CCNR criterion,
and the criterion based on covariance matrices, coincide.
We showed the existence of symmetric bound entangled states,
in particular, a $3\times 3$, five-qubit and six-qubit symmetric
PPT entangled states.

We thank R. Augusiak, A. Doherty, P. Hyllus, T. Moroder, M. Navascues, S. Pironio, R. Werner
and M.M. Wolf for fruitful discussions. We thank especially M.
Lewenstein for many useful discussions on bound entanglement. We
thank the support of the EU (OLAQUI, SCALA, QICS), the National
Research Fund of Hungary OTKA (Contract No. T049234),
%the Hungarian
%Academy of Sciences (J\'anos Bolyai Programme),
the FWF (START
prize) and the Spanish MEC (Ramon y Cajal Programme,
Consolider-Ingenio 2010 project ''QOIT'').


\begin{thebibliography}{99}
\bibitem{basisremark} The
criterion $ 2 \sum |C_{ii}| \leq
[1-\trace(\vr_A^2)]+[1-\trace(\vr_B^2)]$ depends on the choice of
the $M_i$, but it is stronger than the CCNR criterion, if the $M_i$
are chosen appropriately \cite{GH06}.
%Note also that
Filtering algorithms discussed in Ref.~\cite{GH06}
are not guaranteed to converge, as symmetric states are never
of full rank.
%Moreover, a
A symmetric filtering
$\vr \mapsto \FF \otimes \FF \vr (\FF \otimes \FF)^\dagger$
will keep the symmetry of the state, hence the criteria after
filtering will be the same as the PPT criterion.
\bibitem{fannesremark} Already the state in Eq.~(\ref{breuer}) represents a
counterexample as for $\lambda < \tfrac{1}{6}$ it fulfills $\eta \geq 0.$
The term ``symmetric'' in Ref.~\cite{fannes} denotes
the invariant states in our language.
\bibitem{mapremark}
To see this, note that the symmetric space of $N$
qubits is spanned
by the $N+1$ vectors
$
\ket{{0}}_N=\ket{00...00},$
$
\ket{{1}}_N={\rm sym}(\ket{00...01}),
...,
\ket{{N-1}}_N={\rm sym}(\ket{01...11}),
$
$
\ket{{N}}_N=\ket{11...11}.$
In a similar manner, on $n=\tfrac{N}{2}$ qubits, one can define the
$n+1$ vectors $\ket{{0}}_n,\ket{{1}}_n,...,\ket{{n}}_n.$
%$
%\ket{{0}}_n= \ket{00...00},
%$
%$
%\ket{{1}}_n= {\rm sym}(\ket{00...01}),...,
%$
%$
%\ket{{n-1}}_n={\rm sym}(\ket{01...11}),
%$
%$
%\ket{{n}}_n=\ket{11...11}.
%$
Then, the multipartite symmetric space on $N$
qubits  can be viewed as a subspace of the bipartite
symmetric space on this $(n+1)\times (n+1)$
system, as this space is spanned by all
$\ket{i}_n \ket{i}_n$ and $\ket{i}_n \ket{j}_n +\ket{j}_n \ket{i}_n$
and all $\ket{i}_N$ can be written as linear combinations of these.
\bibitem{note} When looking for extensions that are PPT with respect to,
let us say, the $\tfrac{M}{2}:\tfrac{M}{2}$ bipartition,
a further matrix with $(\tfrac{M}{2}+1)^4$ elements has to be stored.
\end{thebibliography}
\end{document}